\documentstyle[sprocl]{article}

\input epsf

\def\be{\begin{equation}}
\def\ee{\end{equation}}
\def\bea{\begin{eqnarray}}
\def\eea{\end{eqnarray}}

\begin{document}

\title{Dual Wilson loop and inter-monopole potential \\
in lattice QCD
\vspace{-0.3 cm}}

\author{A. Tanaka and H. Suganuma}

\address{Research Center for Nuclear Physics, Osaka University,\\ 
Mihogaoka 10-1, Ibaraki 567, 
Japan \\E-mail: atanaka@miho.rcnp.osaka-u.ac.jp} 

\vspace{-0.1cm}

\maketitle\abstracts{We study the dual Wilson loop and
the inter-monopole potential(the 
static potential between the color magnetic monopoles)
in the maximally abelian gauge
to clarify the dual Higgs mechanism induced by monopole condensation.
There is no (color-)electric current in the monopole part,
which includes the essence of the nonperturbative QCD,
and hence the system can be described 
by the dual gauge field $B_\mu$
without the singularity like the Dirac string.
We find that the dual Wilson loop seems to obey the perimeter law,
and the inter-monopole potential becomes 
Yukawa-type in the infrared region.
From the inter-monopole potential,
we estimate the dual gluon mass $m_B$ and the effective size $R$
of the monopole: $m_B \simeq 0.5$GeV, $R \simeq 0.35$fm.
}
\vspace{-0.9cm}
\section{Color Confinement Mechanism and Dual Gauge Formalism}
\vspace{-0.1cm}
We study the confinement mechanism in the
nonperturbative QCD(NP-QCD)
based on the dual superconductor picture.\cite{nan}
In the dual superconductor scenario, 
the QCD vacuum is considered as the dual version of the superconductor,
and the dual Meissner effect, the exclusion of the color-electric
field, is brought by condensation of the 
monopole with the color-magnetic charge instead of the Cooper-pair with 
the electric charge.



As for the appearance of monopoles in QCD,
QCD in the 'tHooft abelian gauge
is reduced into an abelian gauge theory including monopole.
\cite{'tho2,eza,s-s}
In the abelian gauge,
only abelian gluon component
is relevant for NP-QCD, and the abelian gauge
variable can be separated into the photon part
and the monopole part.\cite{}
NP-QCD are realized only in
the monopole part.\cite{s-t,s-s} 

Since the monopole part in QCD dose not include the electric
current $j_\mu$ and only includes the magnetic monopole 
current,
the system resembles the dual version of QED.
Then,  instead of 
the ordinary gauge field $A_\mu$ 
we can introduce the dual gauge field $B_\mu$ without the singularity
like the Dirac string.
The dual gauge field $B_{\mu}$ satisfies 
$\partial_\mu B_\nu - \partial_\nu B_\mu={^*\!F}_{\mu\nu}
= \frac12 \varepsilon_{\mu\nu\rho\sigma}F_{\rho\sigma}$, 
which is the dual version of the ordinary relation,
$F_{\mu\nu} \equiv \partial_\mu A_\nu - \partial_\nu A_\mu$. 
In the dual formalism, the absence of $j_\nu$ is expressed as
the dual Bianchi identity, $j_\nu=\partial_\mu F_{\mu\nu}
={\partial_{\mu}} {^*\!(}\partial \land B)_{\mu\nu}=0$.

In terms of the dual Higgs mechanism, we expect that 
the inter-monopole potential is short-range 
Yukawa-type, and the dual gauge field is massive in the QCD vacuum.
The interaction between the monopoles is described by 
{\it the dual Wilson loop} $W_D$ which is defined 
by the line integral of the dual gauge field $B_\mu$
along a loop $C$:
$W_D(C) \equiv
\exp{i\oint_C B_\mu dx_\mu}=\exp{i\int\!\!\!\int ^*F_{\mu\nu}
d\sigma_{\mu\nu}}$.
The inter-monopole potential is derived from the dual Wilson loop as
$V(R) = -\lim\nolimits_{T \to \infty} \frac{1}{T}\ln W_D(R,T)$.

\vspace{-0.2cm}
\section{Dual Gluon Mass and Monopole Size}
\vspace{-0.1cm}
We study the dual Wilson loop and
the inter-monopole potential in the maximally abelian (MA) gauge
using the SU(2) lattice with $16^4$. 
All measurements are performed at every 100 sweeps
after a thermalization of 5000 sweeps using heat-bath algorithm.
After generating 100 samples of the gauge configuration, 
we evaluate the dual Wilson loop
and the inter-monopole potential.

The dual Wilson loop seems to obey the perimeter law rather
than the area law as shown in Fig.1.
The inter-monopole potential shown in Fig.2 
seems to be short ranged
and flat in comparison with the inter-quark potential.

As for the long-distance region, 
the inter-monopole potential can be fitted  
by the simple Yukawa potential
$V(r) = \frac{-g^2}{4\pi}\frac{e^{-m_Br}}{r}$.
The mass generation of the dual gluon seems to suggest
the dual Higgs mechanism in the NP-QCD vacuum
in the infrared scale.

In the short distance, 
we should consider the size effect of the monopole,
because the QCD-monopole is a soliton like object composed
of gluons.
We introduce the effective size $R$ of the QCD-monopole, and
assume the Gaussian-type distribution of magnetic charge
around it:
\vspace{-0.2cm}
\begin{equation}
\rho({\bf x}) = \frac{1}{(\sqrt{\pi}R)^3}\exp({\frac{-\vert{\bf 
x}\vert^2}{R^2}}).
\label{eq:murnf}
\end{equation}
\vspace{-0.1cm}
Since the system is reduced to be abelian and the
principle of superposition is applicable, 
the Yukawa-type potential with the effective monopole size
is found to be
\vspace{-0.2cm}
\begin{equation}
V({\bf x};R) = \int d^3x_1 \int d^3x_2 \rho({\bf x}_1)\rho({\bf x}_2)
\frac{-g^2}{4\pi}\frac{\exp({-m_B\vert{\bf x-x_1+x_2}\vert})}{
\vert{\bf x-x_1+x_2}\vert}.
\label{eq:murnf}
\end{equation}
\vspace{-0.1cm}

The inter-monopole potential can be fitted by $V({\bf x};R)$ 
as shown in Fig.3 with
the parameters:
the dual gauge coupling $g \simeq 3.1$, the effective monopole size 
$R \simeq 0.35{\rm fm}$
and the dual gluon mass $m_B \simeq 0.5{\rm GeV}$.

\vspace{-0.2cm}
\section{Summary and Concluding Remarks}
\vspace{-0.1cm}
In order to clarify monopole condensation in the QCD vacuum, 
we have calculated the dual Wilson loop, and studied the 
inter-monopole potential in the monopole part 
using the dual gauge formalism based on lattice QCD. 

The dual Wilson loop seems to obey the perimeter law
rather than the area law in the monopole part with $j_\mu=0$.
The inter-monopole potential seems flat and short ranged,
and can be fitted by the simple Yukawa potential in the
long distance.
The Yukawa-type potential with the monopole effective size $R$ seems to
fit the inter-monopole potential both in the whole region of $r$.
The dual gluon mass is estimated as $m_B \simeq {\rm 0.5GeV}$, which is 
consistent with the DGL theory.\cite{s-s}
The generation of the dual gluon mass suggests the 
dual Higgs mechanism in the QCD vacuum.
We find also the monopole size as $R \simeq {\rm 0.35fm}$, which would 
provide the critical scale for NP-QCD
in terms of the dual Higgs theory.
\vspace{-0.5cm}


\begin{figure}[t]
\vspace{-0.5cm}
\hspace{-0.1cm}
\hspace{-0.1cm}
\epsfysize=3.7cm
\epsfbox{dw-a.EPSF}
\epsfysize=3.7cm
\epsfbox{dw-p.EPSF}
\vspace{-0.4cm}
\caption{The lattice QCD data for 
dual Wilson loop $W_D(I,J)$ v.s. its area $(I\times J)$ 
and its perimeter $2(I+J)$ on the SU(2) lattice with $16^4$ and 
$\beta=2.35$.
\label{fig:radish}}
\vspace{0.2cm}
\hspace{-0.1cm}
\epsfysize=3.7cm
\epsfbox{imp-quark.EPSF}
\epsfysize=3.7cm
\epsfbox{imp-y.EPSF}
\vspace{-0.4cm}
\caption{The inter-monopole potential $V(r)$ 
v.s. the distance between the monopoles.
The dashed-dotted
line denotes the linear part of the inter-quark potential in the left figure.
In the right figure, the dashed curve and the solid curve denote
the simple Yukawa potential and the Yukawa-type potential with the 
effective monopole size, respectively. 
\label{fig:radish}}
\vspace{-0.2cm}
\end{figure}
\hspace{0.6cm}



\section*{References}
\begin{thebibliography}{99}
\bibitem{nan}Y. Nambu, {\it Phys. Rev}. {\bf D10} 4262 (1974).



\bibitem{'tho2}G. 't Hooft, {\it Nucl. Phys}. {\bf B190} 455 (1981).

\bibitem{eza}Z. F. Ezawa and A. Iwasaki, {\it Phys. Rev}. {\bf D25} 2681
;{\bf D26} 631(1982).

\bibitem{s-t}H. Suganuma, A. Tanaka, S. Sasaki and O. Miyamura, Proc. of {\em
    Lattice '95}, {\it Nucl. Phys}. {\bf B}(Proc. Suppl.) {\bf 47} 302 (1996).

\bibitem{s-s}H. Suganuma, S. Sasaki and H. Toki, {\it Nucl. Phys}.
{\bf B435} 207 (1995).

\end{thebibliography}
\end{document}